\documentclass[a4paper,11pt]{article}
\pdfoutput=1
\usepackage{jcappub}
\usepackage{hyperref}
\usepackage{amsmath}
\usepackage{breqn}
\usepackage[T1]{fontenc}
\usepackage{hyperref}
\usepackage{xcolor}
\allowdisplaybreaks
\newcommand{\bea}{\begin{eqnarray}}
\newcommand{\eea}{\end{eqnarray}}
\newcommand{\nl}{\nonumber\\&&}
\title{Degenerate higher-order scalar-tensor theories in metric-affine gravity}
\author[a,1]{Hamed Bouzari Nezhad,\note{Independent researcher.}}
\affiliation[a]{Brussels, Belgium}
\emailAdd{hamed.bouzarinezhad@gmail.com}
\abstract{We construct the metric-affine analogue of the quadratic degenerate higher-order scalar-tensor (DHOST) theories. We begin with the metric-affine completion of the quadratic DHOST scalar-tensor action, which is linear in curvature and contains all operators that are at most quadratic in the covariant second derivatives of the scalar field, ensuring that the connection enters only through curvature and these second derivatives. Solving the connection equation by performing a full decomposition of the distortion tensor gives a closed-form effective metric theory. Imposing the standard metric DHOST degeneracy conditions then selects a Palatini Class Ia branch that is fully determined by two free functions in the original action. Analyzing the tensor sector shows that requiring gravitational waves to propagate at the speed of light further restricts the theory to a one-function family. These results provide a detailed and self-contained characterization of the quadratic metric-affine Class Ia sector within this operator basis and identify the theoretical conditions implied by gravitational wave observations.}
\keywords{metric-affine gravity, Palatini scalar-tensor theories, DHOST theories, distortion tensor}
\begin{document}
\maketitle
\flushbottom
\section{Introduction}
\label{Sec: Introduction}
Scalar-tensor theories provide a flexible and powerful framework for exploring departures from General Relativity across cosmological and astrophysical contexts, with applications to dark energy, cosmic acceleration, and screened modified gravity, among many others \cite{Clifton:2011jh,Joyce:2014kja,Koyama:2015vza,Bull:2015stt,Frieman:2008sn,Caldwell:2009ix,Copeland:2006wr,Tsujikawa:2010zza,DeFelice:2010aj,Nojiri:2010wj,Sotiriou:2008rp,Joyce:2016vqv,Ji:2024gdc,Fujii:2003pa,Faraoni:2004pi,Ishak:2018his,Khoury:2003aq,Brax:2012bsa,Burrage:2017qrf,Sakstein:2018fwz,Brax:2021wcv,Kobayashi:2019hrl,Frusciante:2019xia}. A longstanding challenge in constructing such theories is the appearance of Ostrogradski instabilities when higher derivatives of the scalar field are introduced \cite{Ostrogradsky:1850fid,Woodard:2015zca}. Building on the Galileon framework \cite{Nicolis:2008in,Deffayet:2009wt}, the Horndeski class \cite{Horndeski:1974wa,Deffayet:2011gz,Kobayashi:2011nu,Charmousis:2011bf} long represented the most general scalar-tensor theories with second-order field equations, but subsequent developments revealed that consistent higher-order interactions extend beyond this boundary. This insight led to the formulation of the beyond-Horndeski (GLPV) theories \cite{Gleyzes:2014dya,Gleyzes:2014qga} and, more broadly, to the framework of degenerate higher-order scalar-tensor (DHOST) theories \cite{Langlois:2015cwa,Langlois:2015skt,Crisostomi:2016czh,BenAchour:2016fzp,BenAchour:2016cay,Langlois:2018dxi,Deffayet:2015qwa,Lazanu:2024mzj,Heisenberg:2025fxc}. This broader framework was later placed on a systematic footing, with the quadratic and cubic sectors fully classified \cite{Langlois:2015cwa,BenAchour:2016fzp}. Within the quadratic theories, Class Ia plays a central role: it is continuously connected to Horndeski and beyond-Horndeski interactions, stable under disformal transformations \cite{Bettoni:2013diz,Bekenstein:1992pj,Zumalacarregui:2013pma,BenAchour:2016cay}, and includes subclasses with approximate Galileon symmetry that exhibit enhanced non-renormalization properties and controlled radiative corrections \cite{Pirtskhalava:2015nla,Renaux-Petel:2015mga,Heisenberg:2018vsk}. The multi-messenger observation of GW170817 and GRB170817A \cite{LIGOScientific:2017vwq,LIGOScientific:2017zic,LIGOScientific:2018dkp,Goldstein:2017mmi} provided a decisive constraint: the near-luminal propagation of gravitational waves \cite{Creminelli:2017sry,Ezquiaga:2017ekz,Baker:2017hug,Sakstein:2017xjx,Ezquiaga:2018btd}, which restricts modified gravity scenarios to a narrow region of Class Ia \cite{Langlois:2017mxy,Crisostomi:2017lbg,Kase:2018aps,Battye:2018ssx,Noller:2020afd,Hirano:2019scf}. These developments highlight the importance of identifying consistent scalar-tensor interactions in broad geometric settings.

In parallel, the metric-affine formulation of gravity has received renewed attention \cite{Shimada:2018lnm,Iosifidis:2018zwo,BeltranJimenez:2017tkd,Iosifidis:2019dua}. In this approach, the metric and affine connection are independent variables, giving rise to a geometry with torsion, nonmetricity, and the associated distortion tensor. The connection also admits a projective transformation which, in projectively invariant models, can be used to eliminate an otherwise superfluous vector degree of freedom \cite{Shimada:2018lnm,Iosifidis:2018zjj,BeltranJimenez:2020sqf}.\footnote{Projective invariance removes the projective vector mode of the connection in projectively invariant metric-affine actions, but it does not in general guarantee the absence of ghosts. As shown for Ricci-type Palatini theories in \cite{Annala:2022gtl}, additional tensor or vector instabilities may still arise, illustrating that projective symmetry alone is insufficient to ensure stability.} These ideas originate from classic work in differential geometry and gauge approaches to gravity \cite{Schouten:1954,Hehl:1976kj,1976ZNatA..31..111H,Hehl:1994ue,Vitagliano:2010sr} and have been developed in a wide range of modern extensions, including Palatini and hybrid metric-Palatini theories \cite{Sotiriou:2006qn,Olmo:2011uz,Olmo:2012yv,Harko:2011nh,Capozziello:2015lza,Afonso:2018bpv,Barrientos:2018cnx}, Weyl and Weyl-Cartan geometries \cite{BeltranJimenez:2017tkd,BeltranJimenez:2018vdo,Delhom:2019yeo}, and symmetric teleparallel and scalar-nonmetricity models \cite{BeltranJimenez:2017tkd,Jarv:2018bgs,Hohmann:2019nat,Aoki:2018lwx,Aoki:2019rvi,BeltranJimenez:2019esp,Heisenberg:2023lru}. These frameworks have been applied to inflation, compact objects, and non-singular cosmology, and in specific cases offer alternative geometric interpretations of dark energy and dark matter \cite{Afonso:2018bpv,Koivisto:2013kwa,Nezhad:2023dys,Rasanen:2017ivk,Rasanen:2018ihz,Ito:2021ssc,Annala:2020cqj,Rasanen:2022ijc,Jarv:2023sbp,Gialamas:2024jeb}.

Scalar fields have been coupled to metric-affine geometry in several contexts, including Palatini scalar-tensor models \cite{Afonso:2018bpv,Kozak:2018vlp,Galtsov:2018xuc,Rasanen:2017ivk,Rasanen:2018ihz,Ito:2021ssc,BeltranJimenez:2017tkd}, scalar-nonmetricity theories \cite{Jarv:2018bgs,Runkla:2018xrv}, and metric-affine extensions of Galileon and Horndeski interactions \cite{Aoki:2018lwx,Aoki:2019rvi,Helpin:2019vrv}. These constructions demonstrate the richness of scalar-tensor dynamics in the presence of an independent connection, but they typically focus on restricted operator subsets or symmetry-motivated structures. A systematic extension of the DHOST program to the metric-affine setting is still lacking. In particular, the way torsion and nonmetricity modify scalar second-derivative operators and enter the degeneracy conditions has not been explored in full generality. This gap is especially relevant because, in the class of models we consider, the connection equation remains algebraic, ensuring that the independent connection does not introduce additional propagating degrees of freedom prior to degeneracy and making a complete structural analysis tractable. Moreover, identifying the viable higher-derivative sector in the metric-affine framework provides the foundation for systematic studies of cosmology, perturbations, and compact objects in theories sensitive to torsion and nonmetricity \cite{Vitagliano:2010sr,Vitagliano:2013rna,BeltranJimenez:2020sqf,Iosifidis:2020gth,Iosifidis:2018zjj}.

We work within a broad class of Palatini scalar-tensor theories defined by an action that is linear in curvature and contains all operators that are linear and quadratic in the covariant second derivatives of the scalar field. In this framework, the independent connection appears only algebraically, and we determine it exactly by performing a complete tensorial decomposition of the distortion field. Substituting this solution into the action yields an effective metric theory whose coefficients are algebraic functions of the original metric-affine action. Enforcing the standard DHOST degeneracy conditions then isolates a Palatini realization of the quadratic Class Ia family. Although the initial Palatini action contains many independent functions, the degenerate sector reduces to a two-function structure, and the requirement of luminal tensor propagation imposes a further restriction to a one-function theory. This provides a full and consistent characterization of the quadratic Palatini Class Ia sector within our operator basis and sets the stage for phenomenological applications in cosmology and strong gravity regimes.

The remainder of the paper proceeds as follows. Section \ref{Sec: Metric-affine formulation of gravity} reviews the relevant aspects of metric-affine geometry and the decomposition of the independent connection. Section \ref{Sec: Quadratic metric DHOST theories} summarizes the main features of metric DHOST theories and their degeneracy conditions. In Section \ref{Sec: Quadratic Palatini DHOST theories} we present the quadratic metric-affine completion of DHOST theories and the solution to the connection equation of motion. The resulting effective metric theory and its degenerate sector are analyzed in Section \ref{Subsec: Palatini Class Ia DHOST sector}. Section \ref{Subsec: Tensor sector and luminal propagation} examines the tensor sector and identifies the part of the parameter space compatible with luminal gravitational-wave propagation. Additional technical material is collected in the appendices.
\section{Metric-affine formulation of gravity}
\label{Sec: Metric-affine formulation of gravity}
In the metric-affine formulation of gravity \cite{Hehl:1976kj,Iosifidis:2019dua}, the metric $g_{\alpha\beta}$ and the affine connection $\Gamma^{\alpha}{}_{\beta\gamma}$ are treated as independent variables. The affine connection can be decomposed into the Levi-Civita part and the distortion tensor,
\bea
\label{Eq: Connection decomposition}
\Gamma^{\alpha}{}_{\beta\gamma}=\mathring{\Gamma}^{\alpha}{}_{\beta\gamma}+L^{\alpha}{}_{\beta\gamma},
\eea
where $\mathring{}$ denotes quantities constructed from the torsion-free and metric-compatible Levi-Civita connection $\mathring{\Gamma}^{\alpha}{}_{\beta\gamma}$ uniquely defined by the metric. Applying \eqref{Eq: Connection decomposition} to the covariant derivative of the metric gives
\bea
\label{Eq: Cd of metric}
\nabla_{\alpha}g_{\beta\gamma}=\mathring{\nabla}_{\alpha}g_{\beta\gamma}-L_{\beta\alpha\gamma}-L_{\gamma\alpha\beta},
\eea
where the nonmetricity tensor is $Q_{\alpha\beta\gamma}\equiv\nabla_{\alpha}g_{\beta\gamma}=Q_{\alpha(\beta\gamma)}$. Equation \eqref{Eq: Cd of metric} implies that the distortion tensor splits into the disformation $J_{\alpha\beta\gamma}$ and the contortion $K_{\alpha\beta\gamma}$,
\bea
\label{Eq: L split}
L_{\alpha\beta\gamma}=J_{\alpha\beta\gamma}+K_{\alpha\beta\gamma},
\eea
where the two components are determined by nonmetricity and torsion, with torsion defined as $T^{\alpha}{}_{\beta\gamma}=2\Gamma^{\alpha}{}_{[\beta\gamma]}=2L^{\alpha}{}_{[\beta\gamma]}$. Explicitly,
\bea
\label{Eq: J and K}
J_{\alpha\beta\gamma}=\frac{1}{2}(Q_{\alpha\beta\gamma}-Q_{\beta\gamma\alpha}-Q_{\gamma\alpha\beta}),~K_{\alpha\beta\gamma}=\frac{1}{2}(T_{\alpha\beta\gamma}+T_{\beta\alpha\gamma}+T_{\gamma\alpha\beta}),
\eea
where $J_{\alpha\beta\gamma}=J_{\alpha(\beta\gamma)}$ encodes departures from metric compatibility and $K_{\alpha\beta\gamma}=K_{[\alpha|\beta|\gamma]}$ captures torsion. For completeness, we list the standard traces of nonmetricity and torsion,
\bea
\label{Eq: Q and T traces}
Q_{\alpha}=g^{\beta\gamma}Q_{\alpha\beta\gamma},~\hat{Q}_{\alpha}=g^{\beta\gamma}Q_{\beta\alpha\gamma},~T_{\alpha}=g^{\beta\gamma}T_{\beta\alpha\gamma},~\hat{T}^{\alpha}=\frac{1}{6}\epsilon^{\alpha\beta\gamma\delta}T_{\beta\gamma\delta}.
\eea

The Riemann curvature tensor is defined solely from the affine connection,
\bea
\label{Eq: Riemann}
R_{\alpha\beta\gamma}{}^{\delta}=2\partial_{[\beta}\Gamma^{\delta}{}_{\alpha]\gamma}+2\Gamma^{\delta}{}_{[\beta|\epsilon|}\Gamma^{\epsilon}{}_{\alpha]\gamma},
\eea
and is antisymmetric only in the first pair of indices. There are three inequivalent Ricci-type contractions,
\bea
\label{Eq: Ricci Tensors}
R_{\alpha\beta}=R_{\alpha\gamma\beta}{}^{\gamma},~\hat{R}_{\alpha\beta}=g_{\beta\epsilon}g^{\gamma\delta}R_{\alpha\gamma\delta}{}^{\epsilon},~\tilde{R}_{\alpha\beta}=\tilde{R}_{[\alpha\beta]}=R_{\alpha\beta\gamma}{}^{\gamma},
\eea
corresponding to the usual Ricci tensor, the co-Ricci tensor, and the homothetic tensor, respectively. There is only one Ricci scalar,
\bea
\label{Eq: Ricci scalar}
g^{\alpha\beta}R_{\alpha\beta}=R,~g^{\alpha\beta}\hat{R}_{\alpha\beta}=-R,~g^{\alpha\beta}\tilde{R}_{\alpha\beta}=0.
\eea

For later use, we define the Einstein tensor constructed from the metric-affine curvatures,
\bea
\label{Eq: Einstein tensor}
G_{\alpha\beta}=\frac{1}{2}(R_{\alpha\beta}-\hat{R}_{\alpha\beta}-g_{\alpha\beta}R),
\eea
which reduces to the standard Einstein tensor upon imposing the Levi-Civita connection. Using Eqs. \eqref{Eq: Connection decomposition} and \eqref{Eq: Riemann}, the Riemann tensor can be written in terms of the distortion tensor,
\bea
\label{Eq: Riemann in terms of L}
R_{\alpha\beta\gamma}{}^{\delta}=\mathring{R}_{\alpha\beta\gamma}{}^{\delta}+2\mathring{\nabla}_{[\beta}L^{\delta}{}_{\alpha]\gamma}+2L^{\delta}{}_{[\beta|\epsilon|}L^{\epsilon}{}_{\alpha]\gamma}.
\eea

Finally, the connection admits the local projective shift \cite{Hehl:1978zkk},
\bea
\label{Eq: Projective shift}
\Gamma^{\alpha}{}_{\beta\gamma}\longrightarrow\Gamma^{\alpha}{}_{\beta\gamma}+\delta^{\alpha}{}_{\gamma}\xi_{\beta},
\eea
under which the curvature transforms as
\bea
\label{Eq: Riemann projective variation}
R_{\alpha\beta\gamma}{}^{\delta}\longrightarrow R_{\alpha\beta\gamma}{}^{\delta}-2\delta_{\gamma}{}^{\delta}\partial_{[\alpha}\xi_{\beta]},
\eea
showing that the Ricci scalar is invariant under projective transformation of the connection.
\section{Quadratic metric DHOST theories}
\label{Sec: Quadratic metric DHOST theories}
Degenerate higher-order scalar-tensor (DHOST) theories provide the most general metric scalar-tensor models that propagate one scalar degree of freedom in addition to the two tensor modes, while avoiding the Ostrogradski instability through the degeneracy of the Lagrangian \cite{Horndeski:1974wa,Deffayet:2011gz,Gleyzes:2014dya,Crisostomi:2016czh,BenAchour:2016cay,Langlois:2015cwa,Langlois:2018dxi}. Restricting to operators that are at most quadratic in the covariant second derivatives of the scalar field and linear in the curvature, the general metric action takes the form
\bea
\label{Eq: Metric action}
S_{\mathrm{Metric}}&=&\int d^{4}x\sqrt{-g}\Big[f\mathring{R}+\alpha_{1}\mathring{\phi}_{\alpha\beta}\mathring{\phi}^{\alpha\beta}+\alpha_{2}(\mathring{\phi}_{\alpha}{}^{\alpha})^{2}+\alpha_{3}\mathring{\phi}_{\alpha}{}^{\alpha}\mathring{\phi}^{\beta}\mathring{\phi}^{\gamma}\mathring{\phi}_{\beta\gamma}\nl+\alpha_{4}\mathring{\phi}^{\alpha}\mathring{\phi}_{\gamma}\mathring{\phi}_{\alpha\beta}\mathring{\phi}^{\beta\gamma}+\alpha_{5}(\mathring{\phi}^{\alpha}\mathring{\phi}^{\beta}\mathring{\phi}_{\alpha\beta})^{2}\Big],
\eea
where $\mathring{\phi}_{\alpha}=\mathring{\nabla}_{\alpha}\phi$, $\mathring{\phi}_{\alpha\beta}=\mathring{\nabla}_{\alpha}\mathring{\nabla}_{\beta}\phi$, and $\mathring{R}$ is the Ricci scalar constructed from the Levi-Civita connection. The functions $f$ and $\alpha_{i}$ depend on $\phi$ and on $X=g^{\alpha\beta}\mathring{\phi}_{\alpha}\mathring{\phi}_{\beta}$. Up to integration by parts, this basis contains all independent scalars that can be constructed from $g_{\alpha\beta}$, $\phi$, and at most two covariant derivatives of the scalar field.\footnote{In addition to the terms displayed in \eqref{Eq: Metric action}, the most general quadratic DHOST Lagrangian may also include a term depending only on $\phi$ and $X$, operators that are at most linear in $\mathring{\phi}_{\alpha\beta}$, and a coupling between the Einstein tensor and $\mathring{\phi}_{\alpha}\mathring{\phi}_{\beta}$. The zeroth-order and linear second-derivative operators do not modify the DHOST degeneracy conditions and can be omitted, while the Einstein-tensor coupling simply shifts the coefficient of $\mathring{R}$ and reproduces quadratic second-derivative terms already contained in \eqref{Eq: Metric action} \cite{BenAchour:2016cay}.}

Although the action \eqref{Eq: Metric action} contains higher derivatives, the absence of Ostrogradski ghosts follows from the degeneracy of its kinetic matrix. Writing the Lagrangian in general quadratic form and performing the covariant $3+1$ decomposition of \cite{Crisostomi:2016czh}, the determinant of the kinetic matrix can be expressed as
\bea
\label{Eq: Degeneracy I}
D_{0}+D_{1}A_{\ast}^{2}+D_{2}A_{\ast}^{4}=0,
\eea
where $A_{\ast}=n^{\mu}\mathring{\nabla}_{\mu}\phi$ is the normal derivative of the scalar field, and the functions $D_{i}$ depend on $f(\phi,X)$ and on the coefficients $\alpha_{i}(\phi,X)$ multiplying the quadratic invariants. Explicitly \cite{Crisostomi:2016czh},
\bea
\label{Eq: Degeneracy II}
&&D_{0}=-4(\alpha_{1}+\alpha_{2})\big[-2f^{2}-8f_{X}^{2}X^{2}+fX(4f_{X}+2\alpha_{1}+X\alpha_{4})\big],\nl
D_{1}=-8f^{2}(\alpha_{3}+\alpha_{4})+4X\big[16f_{X}^{2}\alpha_{2}-4f_{X}\alpha_{1}(\alpha_{1}+5\alpha_{2}+X\alpha_{3})+\alpha_{1}(\alpha_{1}+3\alpha_{2})\nl\times(2\alpha_{1}+X\alpha_{4})\big]+f\big[48f_{X}^{2}-4\alpha_{1}^{2}-16\alpha_{1}\alpha_{2}+12X\alpha_{1}\alpha_{3}-X^{2}\alpha_{3}^{2}+8f_{X}(-2\alpha_{1}\nl+4\alpha_{2}+X\alpha_{3})-16X\alpha_{2}\alpha_{4}+4X^{2}(\alpha_{1}+\alpha_{2})\alpha_{5}\big],\nl
D_{2}=4\alpha_{1}^{3}+16f_{X}^{2}(\alpha_{1}+2\alpha_{2})-4fX\alpha_{3}^{2}-8f_{X}\big[2f\alpha_{3}+\alpha_{1}(2\alpha_{1}+4\alpha_{2}-X\alpha_{3})\big]\nl+8f(f+2X\alpha_{2})\alpha_{5}+\alpha_{1}^{2}\big[8\alpha_{2}-4X(\alpha_{3}+X\alpha_{5})\big]+\alpha_{1}\big[8f\alpha_{3}+3X^{2}(\alpha_{3}^{2}-4\alpha_{2}\alpha_{5})\big].
\eea

The degeneracy conditions $D_{0}=D_{1}=D_{2}=0$ ensure that the kinetic matrix has vanishing determinant and thus generate a primary constraint that removes the Ostrogradski mode. Solving these relations among the functions $\alpha_{i}(\phi,X)$ yields the complete classification of quadratic DHOST theories into Classes I, II, and III, with Class Ia being continuously connected to Horndeski and beyond-Horndeski models \cite{Langlois:2015cwa,Langlois:2018dxi}. Observational constraints on the near-luminal speed of gravitational waves select viable subclasses of Class Ia with $c_{T}^{2}\simeq1$ \cite{Creminelli:2017sry,Ezquiaga:2017ekz}. The action \eqref{Eq: Metric action}, together with the degeneracy conditions above, therefore provides the complete and ghost-free quadratic metric DHOST framework that will serve as a reference for the metric-affine construction developed below.
\section{Quadratic Palatini DHOST theories}
\label{Sec: Quadratic Palatini DHOST theories}
We now turn to the metric-affine formulation and construct the Palatini counterparts of the quadratic DHOST theories reviewed in Section \ref{Sec: Quadratic metric DHOST theories}. In this framework, the corresponding action reads
\bea
\label{Eq: Palatini action}
S_{\mathrm{M\text{-}A}}&=&\int d^{4}x\sqrt{-g}
\Big[F_{1}R+F_{2}G_{\alpha\beta}\phi^{\alpha}\phi^{\beta}+P+Q_{1}\phi_{\alpha}{}^{\alpha}+Q_{2}\phi_{\alpha\beta}\phi^{\alpha}\phi^{\beta}+A_{1}\phi_{\alpha\beta}\phi^{\alpha\beta}\nl+A_{2}(\phi_{\alpha}{}^{\alpha})^{2}+A_{3}\phi_{\alpha}{}^{\alpha}\phi^{\beta}\phi^{\gamma}\phi_{\beta\gamma}+A_{4}\phi^{\alpha}\phi_{\gamma}\phi_{\alpha\beta}\phi^{\beta\gamma}+A_{5}(\phi^{\alpha}\phi^{\beta}\phi_{\alpha\beta})^{2}\Big],
\eea
where $\phi_{\alpha}=\nabla_{\alpha}\phi$, $\phi_{\alpha\beta}=\nabla_{\alpha}\nabla_{\beta}\phi$, $R$ is the Palatini Ricci scalar, and $G_{\alpha\beta}$ is the Palatini Einstein tensor in Eq. \eqref{Eq: Einstein tensor}.\footnote{Because $\phi$ is a scalar, $\phi_{\alpha}=\nabla_{\alpha}\phi=\mathring{\nabla}_{\alpha}\phi$, so $X=g^{\alpha\beta}\phi_{\alpha}\phi_{\beta}$ is the same in both formulations. We write $\mathring{\phi}_{\alpha}$ only in the metric action and use $\phi_{\alpha}$ in the metric-affine analysis to avoid superfluous Levi-Civita notation.} The functions $F_{1}$, $F_{2}$, $P$, $Q_{1}$, $Q_{2}$, and $A_{i}$ depend on $\phi$ and $X$. Compared to the purely metric case, we retain all terms proportional to $P$, $Q_{1}$, $Q_{2}$, and $F_{2}$, since $\phi_{\alpha\beta}$ and the Palatini curvature tensors depend on the distortion tensor and these contributions enter the algebraic distortion tensor equation of motion. The departure from the metric second derivative arises entirely from the distortion tensor,
\bea
\label{Eq: Departure}
\phi_{\alpha\beta}=\mathring{\phi}_{\alpha\beta}-L^{\gamma}{}_{\alpha\beta}\phi_{\gamma}.
\eea

Before proceeding, a few clarifying remarks on the structure of the Palatini action \eqref{Eq: Palatini action} are in order. We define $\phi_{\alpha\beta}\equiv\nabla_{\alpha}\nabla_{\beta}\phi$, where $\nabla_{\alpha}$ denotes the covariant derivative associated with the independent metric-affine connection. In general, $\phi_{\alpha\beta}\neq\phi_{\beta\alpha}$, with the antisymmetric part entirely determined by torsion, according to $2\phi_{[\alpha\beta]}=-T^{\gamma}{}_{\alpha\beta}\nabla_{\gamma}\phi$. Treating $\phi_{\alpha\beta}\phi^{\alpha\beta}$ and $\phi_{\alpha\beta}\phi^{\beta\alpha}$ as independent operators would therefore introduce an explicit scalar-torsion invariant proportional to $(T\cdot\nabla\phi)^2$. Since our aim is to construct the Palatini completion of the quadratic DHOST operator basis, we retain a single contraction of $\phi_{\alpha\beta}$ at quadratic order, with torsion entering only implicitly through the independent connection. The same remark applies to the remaining quadratic operators in \eqref{Eq: Palatini action}: alternative index orderings would likewise generate explicit scalar-torsion invariants and are not included. Alternative definitions such as $\phi_{\alpha}{}^{\beta}=\nabla_{\alpha}(g^{\beta\gamma}\nabla_{\gamma}\phi)$ differ from $g^{\beta\gamma}\nabla_{\alpha}\nabla_{\gamma}\phi$ by terms proportional to $\nabla_{\alpha}g^{\beta\gamma}$, i.e.\ explicit scalar-nonmetricity contributions, and would correspond to a broader class of theories than the one considered here.\footnote{In this work, the metric-affine completion of the quadratic metric DHOST action refers to the promotion of the complete quadratic scalar-tensor operator basis of the metric DHOST theory to a framework with an independent affine connection, including operators whose role in the degeneracy conditions is trivial in the metric formulation but becomes non-trivial once the connection is treated as an independent variable. Independent scalar-torsion or scalar-nonmetricity invariants are not included.}

The action \eqref{Eq: Palatini action} is not projectively invariant, since $\phi_{\alpha\beta}$ transforms non-trivially under the projective shift \eqref{Eq: Projective shift}. Nevertheless, the equation of motion for the distortion tensor is purely algebraic, so $L^{\alpha}{}_{\beta\gamma}$ can be consistently integrated out.

Using the identity \eqref{Eq: Departure} together with the decomposition of the Palatini curvature in terms of the distortion tensor, Eq. \eqref{Eq: Riemann in terms of L}, the variation of \eqref{Eq: Palatini action} with respect to $L^{\alpha}{}_{\beta\gamma}$ yields an equation of the form\footnote{This follows after integrating by parts the $\mathring{\nabla}L$ terms in the curvature decomposition and using the antisymmetrized commutation identity $2\mathring{\nabla}_{[\alpha}\mathring{\nabla}_{\beta]}\mathring{\phi}^{\beta}=-\mathring{R}_{\alpha\beta}\mathring{\phi}^{\beta}$ to eliminate the Ricci term generated by the Einstein tensor coupling.}
\bea
\label{Eq: LEoM}
\mathcal{M}_{\alpha\beta\gamma}{}^{\delta\epsilon\zeta}L_{\delta\epsilon\zeta}=\mathcal{S}_{\alpha\beta\gamma},
\eea
in which
\bea
\label{Eq: S}
\mathcal{S}_{\alpha\beta\gamma}&=&2Q_{2}\mathring{\phi}_{\alpha}\mathring{\phi}_{\beta}\mathring{\phi}_{\gamma}+F_{2}\mathring{\phi}_{\gamma}\mathring{\phi}_{\alpha\beta}+2F_{2,X}\mathring{\phi}_{\beta}\mathring{\phi}_{\gamma}\mathring{\phi}^{\delta}\mathring{\phi}_{\alpha\delta}+4A_{1}\mathring{\phi}_{\alpha}\mathring{\phi}_{\beta\gamma}-F_{2}\mathring{\phi}_{\alpha}\mathring{\phi}_{\beta\gamma}\nl+2A_{4}\mathring{\phi}_{\alpha}\mathring{\phi}_{\gamma}\mathring{\phi}^{\delta}\mathring{\phi}_{\beta\delta}+2A_{4}\mathring{\phi}_{\alpha}\mathring{\phi}_{\beta}\mathring{\phi}^{\delta}\mathring{\phi}_{\gamma\delta}-2F_{2,X}\mathring{\phi}_{\alpha}\mathring{\phi}_{\beta}\mathring{\phi}^{\delta}\mathring{\phi}_{\gamma\delta}+4A_{5}\mathring{\phi}_{\alpha}\mathring{\phi}_{\beta}\mathring{\phi}_{\gamma}\mathring{\phi}^{\delta}\mathring{\phi}^{\epsilon}\mathring{\phi}_{\delta\epsilon}\nl+2A_{3}\mathring{\phi}_{\alpha}\mathring{\phi}_{\beta}\mathring{\phi}_{\gamma}\mathring{\phi}_{\delta}{}^{\delta}-g_{\alpha\beta}\big[(4F_{1,X}-F_{2}-2F_{2,X}X)\mathring{\phi}^{\delta}\mathring{\phi}_{\delta}{}^{\delta}_{\gamma\delta}+\mathring{\phi}_{\gamma}(2F_{1,\phi}+2F_{2,X}\mathring{\phi}^{\delta}\mathring{\phi}^{\epsilon}\mathring{\phi}_{\delta\epsilon}\nl+F_{2}\mathring{\phi}_{\delta}{}^{\delta})\big]+g_{\beta\gamma}\bigg((4F_{1,X}-F_{2}-2F_{2,X}X)\mathring{\phi}^{\delta}\mathring{\phi}_{\alpha\delta}+\mathring{\phi}_{\alpha}\Big\{2\big[F_{1,\phi}+Q_{1}+(A_{3}\nl+F_{2,X})\mathring{\phi}^{\delta}\mathring{\phi}^{\epsilon}\mathring{\phi}_{\delta\epsilon}\big]+(4A_{2}+F_{2})\mathring{\phi}_{\delta}{}^{\delta}\Big\}\bigg),
\eea
and
\bea
\label{Eq: M}
\mathcal{M}_{\alpha\beta\gamma}{}^{\delta\epsilon\zeta}&=&2F_{1}\delta_{\gamma}{}^{\delta}g_{\alpha\beta}g^{\epsilon\zeta}-F_{2}X\delta_{\gamma}{}^{\delta}g_{\alpha\beta}g^{\epsilon\zeta}+F_{2}\delta_{\gamma}{}^{\delta}g^{\epsilon\zeta}\mathring{\phi}_{\alpha}\mathring{\phi}_{\beta}+4A_{1}\delta_{\beta}{}^{\epsilon}\delta_{\gamma}{}^{\zeta}\mathring{\phi}_{\alpha}\mathring{\phi}^{\delta}\nl+4A_{2}g_{\beta\gamma}g^{\epsilon\zeta}\mathring{\phi}_{\alpha}\mathring{\phi}^{\delta}+2A_{3}g^{\epsilon\zeta}\mathring{\phi}_{\alpha}\mathring{\phi}_{\beta}\mathring{\phi}_{\gamma}\mathring{\phi}^{\delta}-F_{2}\delta_{\beta}{}^{\zeta}\delta_{\gamma}{}^{\delta}\mathring{\phi}_{\alpha}\mathring{\phi}^{\epsilon}+2A_{4}\delta_{\beta}{}^{\zeta}\mathring{\phi}_{\alpha}\mathring{\phi}_{\gamma}\mathring{\phi}^{\delta}\mathring{\phi}^{\epsilon}\nl+\delta_{\alpha}{}^{\zeta}\Big\{F_{2}\mathring{\phi}_{\beta}(g^{\delta\epsilon}\mathring{\phi}_{\gamma}-\delta_{\gamma}{}^{\epsilon}\mathring{\phi}^{\delta})+\delta_{\beta}{}^{\delta}\big[(-2F_{1}+F_{2}X)\delta_{\gamma}{}^{\epsilon}-F_{2}\mathring{\phi}_{\gamma}\mathring{\phi}^{\epsilon}\big]+g_{\beta\gamma}\big[(2F_{1}\nl-F_{2}X)g^{\delta\epsilon}+F_{2}\mathring{\phi}^{\delta}\mathring{\phi}^{\epsilon}\big]\Big\}+2A_{4}\delta_{\gamma}{}^{\epsilon}\mathring{\phi}_{\alpha}\mathring{\phi}_{\beta}\mathring{\phi}^{\delta}\mathring{\phi}^{\zeta}+F_{2}\delta_{\gamma}{}^{\delta}g_{\alpha\beta}\mathring{\phi}^{\epsilon}\mathring{\phi}^{\zeta}+2A_{3}g_{\beta\gamma}\mathring{\phi}_{\alpha}\mathring{\phi}^{\delta}\mathring{\phi}^{\epsilon}\mathring{\phi}^{\zeta}\nl+4A_{5}\mathring{\phi}_{\alpha}\mathring{\phi}_{\beta}\mathring{\phi}_{\gamma}\mathring{\phi}^{\delta}\mathring{\phi}^{\epsilon}\mathring{\phi}^{\zeta}+\delta_{\alpha}{}^{\epsilon}\delta_{\gamma}{}^{\delta}\big[(-2F_{1}+F_{2}X)\delta_{\beta}{}^{\zeta}-F_{2}\mathring{\phi}_{\beta}\mathring{\phi}^{\zeta}\big],
\eea
where $\mathcal{M}_{\alpha\beta\gamma}{}^{\delta\epsilon\zeta}$ depends on $g_{\alpha\beta}$, $\phi$ and its first derivatives. In the following we restrict to the generic branch in which $\mathcal{M}_{\alpha\beta\gamma}{}^{\delta\epsilon\zeta}$ is invertible, so that Eq. \eqref{Eq: LEoM} admits a unique regular solution for $L_{\alpha\beta\gamma}$. Eq. \eqref{Eq: LEoM} can be solved by the following ansatz
\bea
\label{Eq: Ansatz}
L_{\alpha\beta\gamma}&=&l_{1}g_{\beta\gamma}\mathring{\phi}_{\alpha}+l_{2}g_{\alpha\gamma}\mathring{\phi}_{\beta}+l_{3}g_{\alpha\beta}\mathring{\phi}_{\gamma}+l_{4}\mathring{\phi}_{\alpha}\mathring{\phi}_{\beta}\mathring{\phi}_{\gamma}+l_{5}\mathring{\phi}_{\gamma}\mathring{\phi}_{\alpha\beta}+l_{6}\mathring{\phi}_{\beta}\mathring{\phi}_{\alpha\gamma}+l_{7}\mathring{\phi}_{\alpha}\mathring{\phi}_{\beta\gamma}\nl+l_{8}g_{\beta\gamma}\mathring{\phi}^{\delta}\mathring{\phi}_{\alpha\delta}+l_{9}\mathring{\phi}_{\beta}\mathring{\phi}_{\gamma}\mathring{\phi}^{\delta}\mathring{\phi}_{\alpha\delta}+l_{10}g_{\alpha\gamma}\mathring{\phi}^{\delta}\mathring{\phi}_{\beta\delta}+l_{11}\mathring{\phi}_{\alpha}\mathring{\phi}_{\gamma}\mathring{\phi}^{\delta}\mathring{\phi}_{\beta\delta}\nl+l_{12}g_{\alpha\beta}\mathring{\phi}^{\delta}\mathring{\phi}_{\gamma\delta}+l_{13}\mathring{\phi}_{\alpha}\mathring{\phi}_{\beta}\mathring{\phi}^{\delta}\mathring{\phi}_{\gamma\delta}+l_{14}g_{\beta\gamma}\mathring{\phi}_{\alpha}\mathring{\phi}_{\delta}{}^{\delta}+l_{15}g_{\alpha\gamma}\mathring{\phi}_{\beta}\mathring{\phi}_{\delta}{}^{\delta}+l_{16}g_{\alpha\beta}\mathring{\phi}_{\gamma}\mathring{\phi}_{\delta}{}^{\delta}\nl+l_{17}\mathring{\phi}_{\alpha}\mathring{\phi}_{\beta}\mathring{\phi}_{\gamma}\mathring{\phi}_{\delta}{}^{\delta}+l_{18}g_{\beta\gamma}\mathring{\phi}_{\alpha}\mathring{\phi}^{\delta}\mathring{\phi}^{\epsilon}\mathring{\phi}_{\delta\epsilon}+l_{19}g_{\alpha\gamma}\mathring{\phi}_{\beta}\mathring{\phi}^{\delta}\mathring{\phi}^{\epsilon}\mathring{\phi}_{\delta\epsilon}+l_{20}g_{\alpha\beta}\mathring{\phi}_{\gamma}\mathring{\phi}^{\delta}\mathring{\phi}^{\epsilon}\mathring{\phi}_{\delta\epsilon}\nl+l_{21}\mathring{\phi}_{\alpha}\mathring{\phi}_{\beta}\mathring{\phi}_{\gamma}\mathring{\phi}^{\delta}\mathring{\phi}^{\epsilon}\mathring{\phi}_{\delta\epsilon},
\eea
where $l_{i}=l_{i}(\phi,X)$. Substituting this ansatz into Eq. \eqref{Eq: LEoM} allows one to solve explicitly for the coefficients $l_{i}$, whose expressions are listed in appendix \ref{App: Distortion tensor coefficients}. Inserting the resulting distortion tensor into the action \eqref{Eq: Palatini action}, we obtain
\bea
\label{Eq: Effective action}
S_{\mathrm{Eff}}&=&\int d^{4}x\sqrt{-g}\Big[\tilde{f}\mathring{R}+\tilde{P}+\tilde{Q}_{1}\mathring{\phi}_{\alpha}{}^{\alpha}+\tilde{Q}_{2}\mathring{\phi}_{\alpha\beta}\mathring{\phi}^{\alpha}\mathring{\phi}^{\beta}+\tilde{A}_{1}\mathring{\phi}_{\alpha\beta}\mathring{\phi}^{\alpha\beta}+\tilde{A}_{2}(\mathring{\phi}_{\alpha}{}^{\alpha})^{2}\nl+\tilde{A}_{3}\mathring{\phi}_{\alpha}{}^{\alpha}\mathring{\phi}^{\beta}\mathring{\phi}^{\gamma}\mathring{\phi}_{\beta\gamma}+\tilde{A}_{4}\mathring{\phi}^{\alpha}\mathring{\phi}_{\gamma}\mathring{\phi}_{\alpha\beta}\mathring{\phi}^{\beta\gamma}+\tilde{A}_{5}(\mathring{\phi}^{\alpha}\mathring{\phi}^{\beta}\mathring{\phi}_{\alpha\beta})^{2}\Big],
\eea
where $\tilde{f}$, $\tilde{P}$, $\tilde{Q}_{1}$, $\tilde{Q}_{2}$, and $\tilde{A}_{i}$ have been given in appendix \ref{App: Effective action coefficients}. A key structural result is that, although the functions $Q_{1}$ and $Q_{2}$ enter the algebraic solution for the distortion tensor through the coefficients $l_{i}$ (see appendix \ref{App: Distortion tensor coefficients}), their role is entirely confined to the functions $\tilde{P}$, $\tilde{Q}_{1}$, and $\tilde{Q}_{2}$ in the effective metric action. Crucially, they do not induce any contributions to the quadratic coefficients $\tilde{A}_{i}$, which exhaust the second-derivative structure relevant for the DHOST analysis. Hence the quadratic sector obtained after integrating out the connection is fully controlled by $F_{1}$, $F_{2}$, and the original quadratic operators, independently of $P$, $Q_{1}$, and $Q_{2}$.
\subsection{Palatini Class Ia DHOST sector}
\label{Subsec: Palatini Class Ia DHOST sector}
We now impose the metric degeneracy conditions on the effective action \eqref{Eq: Effective action} in order to isolate the Palatini analogue of the Class Ia DHOST subclass. The first condition, $\tilde{A}_{1}+\tilde{A}_{2}=0$, fixes one algebraic combination of the Palatini coefficients and allows us to solve for $A_{2}$ in terms of $F_{1}$, $F_{2}$, $A_{1}$, $A_{3}$, $A_{4}$, and $A_{5}$. Substituting this result into the second degeneracy condition yields an explicit expression for $A_{4}$, with two algebraic branches $A_{4}^{(\pm)}$. However, once integrated out, both branches lead to the same regular metric coefficients; hence, they correspond to the same physical theory. The third degeneracy condition depends only on the remaining function $A_{1}$. Solving for $A_{1}$ again produces two branches, but one of them renders the effective combinations $\tilde{A}_{4}$ and $\tilde{A}_{5}$ indeterminate ($0/0$). This branch therefore fails to produce a consistent quadratic action and is discarded. The remaining solution yields a well-defined, non-singular effective theory. We also explicitly verify that it satisfies $\tilde{f}-X\tilde{A}_{1}\neq0$, which confirms that we are selecting the non-minimal Class Ia branch rather than the alternative $\tilde{f}-X\tilde{A}_{1}=0$ degeneracy.

A notable outcome of this analysis is that, after substituting the solutions for $A_{1}$, $A_{2}$, and $A_{4}$, the entire Class Ia sector is determined by the two free functions $F_{1}(\phi,X)$ and $F_{2}(\phi,X)$ alone. In particular, although $A_{3}$ and $A_{5}$ appear as free functions at the level of the Palatini action and enter the intermediate effective coefficients, their functional dependence cancels identically once the degeneracy conditions are imposed; consequently, they do not label independent degenerate theories. As a consequence, all quadratic Palatini coefficients $A_{i}$ become algebraic functions of $F_{1}$ and $F_{2}$, and the resulting metric action depends only on these two functions.

The Class Ia Palatini DHOST action therefore reduces to a two-function family,
\bea
\label{Eq: Class Ia}
S_{\mathrm{Ia}}&=&\int d^{4}x\sqrt{-g}\Big\{\tilde{f}\mathring{R}+\tilde{p}+\tilde{q}_{1}\mathring{\phi}_{\alpha}{}^{\alpha}+\tilde{q}_{2}\mathring{\phi}_{\alpha\beta}\mathring{\phi}^{\alpha}\mathring{\phi}^{\beta}+\tilde{a}_{1}\big[\mathring{\phi}_{\alpha\beta}\mathring{\phi}^{\alpha\beta}-(\mathring{\phi}_{\alpha}{}^{\alpha})^{2}\big]+\nl+\tilde{a}_{3}\mathring{\phi}_{\alpha}{}^{\alpha}\mathring{\phi}^{\beta}\mathring{\phi}^{\gamma}\mathring{\phi}_{\beta\gamma}+\tilde{a}_{4}\mathring{\phi}^{\alpha}\mathring{\phi}_{\gamma}\mathring{\phi}_{\alpha\beta}\mathring{\phi}^{\beta\gamma}+\tilde{a}_{5}(\mathring{\phi}^{\alpha}\mathring{\phi}^{\beta}\mathring{\phi}_{\alpha\beta})^{2}\Big\}
\eea
where $\tilde{p}$, $\tilde{q}_{1}$, $\tilde{q}_{2}$, $\tilde{a}_{1}$, $\tilde{a}_{2}$, $\tilde{a}_{3}$, $\tilde{a}_{4}$, and $\tilde{a}_{5}$ for $P=Q_{1}=Q_{2}=0$ are listed in appendix \ref{App: Class Ia coefficients}. Eq. \eqref{Eq: Class Ia} satisfies all three degeneracy conditions by construction.

Although the operator structure of the effective action \eqref{Eq: Class Ia} coincides with that of a generic quadratic DHOST model in the Class Ia branch, the corresponding coefficients are not independent. In our Palatini construction they are all generated algebraically from only two functions, namely $F_{1}$ and $F_{2}$ appearing in the original action. The combinations $\tilde{f}$, $\tilde{p}$, $\tilde{q}_{i}$ and $\tilde{a}_{i}$ in \eqref{Eq: Class Ia} therefore span a two-function subset of the full Class Ia family, and their explicit expressions in terms of $F_{1}$, $F_{2}$, and their derivatives are given in appendix \ref{App: Class Ia coefficients}. Once the Palatini couplings are specified, all metric DHOST coefficients in \eqref{Eq: Class Ia} are fixed, defining the characteristic Palatini slice of Class Ia theories.
\subsection{Tensor sector and luminal propagation}
\label{Subsec: Tensor sector and luminal propagation}
The propagation speed of gravitational waves follows from the quadratic action for the tensor perturbations,
\bea
\label{Eq: Quadratic action}
S_{T}^{(2)}=\frac{1}{8}\int dtd^{3}xa^{3}\left[\mathcal{G}_{T}\dot h_{ij}^{2}-\mathcal{F}_{T}\frac{(\partial_{k} h_{ij})^{2}}{a^{2}}\right],
\eea
where $c_{T}^{2}=\mathcal{F}_{T}/\mathcal{G}_{T}$ \cite{Kobayashi:2019hrl,Langlois:2017mxy,deRham:2016wji}. For quadratic DHOST theories in the Class Ia branch one has \cite{Mironov:2020pqh}
\bea
\label{Eq: Definitions}
\mathcal{G}_{T}=2(\tilde{f} - X\tilde{a}_{1}),~\mathcal{F}_{T}=2\tilde{f},
\eea
so that
\bea
\label{Eq: c_{T}^{2} I}
c_{T}^{2} =\frac{\tilde{f}}{\tilde{f}-X\tilde{a}_{1}}.
\eea

Imposing exact luminal propagation $c_{T}^{2}=1$ for scalar-tensor backgrounds with $X\neq0$ immediately yields $\tilde{a}_{1}(\phi,X)=0$. Using the explicit Class Ia expression for $\tilde{a}_{1}$ in terms of the original Palatini functions $F_{1}$ and $F_{2}$
\bea
\label{Eq: Luminal condition}
\tilde{a}_{1}=-F_{2}+\frac{F_{1}}{X}+\frac{F_{1}^{2}F_{2}}{2(F_{1}^{2}+3F_{1,X}^{2}X^{2})}=0.
\eea

The condition \eqref{Eq: Luminal condition} gives the following algebraic relation between $F_{1}$ and $F_{2}$
\bea
\label{Eq: F_{2} luminal}
F_{2}=F_{1}\left(\frac{2}{X}-\frac{6F_{1,X}^{2}X}{F_{1}^{2}+6F_{1,X}^{2}X^{2}}\right).
\eea

Thus, demanding $c_{T}^{2}=1$ for generic scalar-tensor configurations reduces the two-function Palatini Class Ia family to a one-function subclass specified entirely by $F_{1}$, with $F_{2}$ fixed by \eqref{Eq: F_{2} luminal}. In addition to $c_{T}^{2}=1$, the absence of ghost and gradient instabilities requires $\mathcal{G}>0$ and $\mathcal{F}>0$, which in the luminal branch simply reduce to $\tilde{f}>0$. This constitutes the exact-luminal Palatini DHOST Class Ia sector.

\section{Conclusions}
\label{Sec: Conclusions}
In this work we have constructed the complete quadratic degenerate higher-order scalar-tensor sector in the metric-affine framework, where the metric $g_{\alpha\beta}$ and the affine connection $\Gamma^{\alpha}{}_{\beta\gamma}$ are independent variables. For comparison, in the purely metric formulation the quadratic DHOST action \eqref{Eq: Metric action} reduces, under the Class Ia degeneracy conditions, to a three-function family specified by $f$, $\alpha_{2}$, and $\alpha_{3}$, with all remaining coefficients fixed algebraically by the degeneracy relations.

In the Palatini setting we begin from a substantially larger operator basis: the quadratic scalar-tensor action, obtained as the metric-affine completion of the metric DHOST theories, includes all scalar invariants built from first and second derivatives of the scalar field, the Palatini Ricci scalar, the Einstein-tensor coupling, and the dependence on torsion and nonmetricity encoded in the distortion tensor. Solving the algebraic connection equation using a systematic $21$-component decomposition of the distortion field yields an exact closed-form expression for the effective metric theory. Imposing the standard metric DHOST degeneracy conditions on this effective action collapses the entire Palatini quadratic Lagrangian to a two-function family determined solely by $F_{1}$ and $F_{2}$. All quadratic coefficients $A_{i}$ become algebraic functions of these two, providing the Palatini analogue of metric Class Ia. Thus, the independent connection does not enlarge the space of viable DHOST interactions; rather, it leads to a more constrained structure, reducing the number of free functions from three to two.

These results can be placed in the context of previous works on scalar-tensor and curvature-based extensions of gravity in metric-affine geometry. Earlier analyses of Palatini scalar-tensor and Ricci-based theories \cite{Sotiriou:2006qn,Olmo:2011uz,Harko:2011nh,Capozziello:2015lza,Afonso:2017aci,Barrientos:2018cnx,Jarv:2018bgs}, derivative couplings and scalar-nonmetricity models \cite{Jarv:2018bgs,Galtsov:2018xuc,Runkla:2018xrv}, and metric-affine versions of selected Horndeski and Galileon interactions \cite{Aoki:2018lwx,Aoki:2019rvi} demonstrate that torsion, nonmetricity, and an independent connection can influence scalar dynamics. However, these studies typically focus on restricted operator subsets or symmetry-motivated constructions, and do not perform a DHOST degeneracy analysis of the full quadratic scalar-tensor sector.

Quadratic torsion and nonmetricity invariants have also been systematically explored in general metric-affine gravity, from early gauge-theoretic formulations \cite{Hehl:1994ue} to recent analyses of quadratic MAG theories that explicitly solve the connection equations and investigate their dynamics \cite{Vitagliano:2010sr,Vitagliano:2013rna,Iosifidis:2018zwo,BeltranJimenez:2019acz,Iosifidis:2021tvx,Iosifidis:2021bad,Iosifidis:2021fnq,Iosifidis:2024bsq,Katsoulas:2025mcu}. While these works classify broad families of quadratic distortion invariants and study their phenomenology, they do not couple the connection to a scalar field in the DHOST framework or impose degeneracy conditions that eliminate the Ostrogradski mode. By contrast, our analysis begins with the quadratic Palatini scalar-tensor action obtained as the metric-affine completion of quadratic metric DHOST theories, solves the connection equation of motion exactly, and enforces the DHOST degeneracy relations, thereby identifying the complete degenerate Palatini sector.

Unlike projectively invariant scalar-tensor constructions \cite{Aoki:2018lwx,Aoki:2019rvi}, the action considered here is not projectively invariant, since $\phi_{\alpha\beta}=\nabla_\alpha\nabla_\beta\phi$ transforms nontrivially under projective shifts. This does not pose a difficulty: the connection equation is purely algebraic, so the projective mode does not propagate and can be integrated out. In our framework it is the DHOST degeneracy conditions, not projective symmetry, that ensure the correct number of propagating degrees of freedom. This establishes a clear correspondence between the metric and Palatini formulations and explains how the complete Palatini analogue of Class Ia emerges.

A natural direction for future work is to investigate the phenomenology of the Palatini DHOST sector identified here. The reduced two-function structure (and its one-function luminal subset) provides a controlled setting in which to analyze cosmology, screening mechanisms, and strong gravity solutions, and to assess the observational viability of scalar-tensor theories with an independent connection. Further developments include extending the present framework to cubic and higher-order interactions, examining whether projectively invariant realizations exist within the degenerate sector, and exploring possible links with quadratic metric-affine gravity and other non-Riemannian models. The program may also be applied to Palatini counterparts of other DHOST classes. Such investigations would clarify the broader landscape of degenerate scalar-tensor theories in metric-affine geometry and may reveal deeper correspondences between metric and Palatini formulations beyond the quadratic level.
\appendix
\section{Distortion tensor coefficients}
\label{App: Distortion tensor coefficients}

In this appendix we present the explicit expressions for the distortion tensor coefficients appearing in Eq. \eqref{Eq: Ansatz}. Several of these coefficients satisfy simple algebraic identities, which streamline their presentation,
\bea
\label{Eq: Relations between ls}
&&l_{3}=-l_{1},\nl
l_{6}=\frac{2A_{1}}{2A_{1}-F_{2}}l_{5},\nl
l_{7}=-l_{5},\nl
l_{12}=-l_{8},\nl
l_{13}=-l_{9}-\frac{2l_{6}}{X},\nl
l_{16}=-l_{14},\nl
l_{18}=-l_{20}.
\eea

In addition, several recurrent structures appear in the denominators of the $l_{i}$ coefficients, and isolating these allows the following expressions to be written in a more compact form
\bea
\label{Eq: Action shorthands}
&&\tilde{f}=F_{1}-\frac{1}{2}F_{2}X,\nl
d_{1}=2(\tilde{f}+A_{1}X),\nl
d_{2}=-4A_{1}(A_{1}+4A_{2})X-4\big[3A_{2}A_{4}+A_{1}(A_{3}+A_{4})\big]X^{2}+\big[3A_{3}^{2}-4(A_{1}+3A_{2})A_{5}\big] X^{3}\nl\phantom{d_{2}=}+8\tilde{f}\big[A_{1}+A_{2}+X(A_{3}+A_{4}+A_{5}X)\big],\nl
d_{3}=16A_{1}F_{1}^{2}+2\tilde{f}X(4A_{1}^{2}-A_{4}^{2}X^{2}).
\eea

We will work in the generic case where $d_{1}$, $d_{2}$ and $d_{3}$ do not vanish, so that the algebraic connection equation admits a regular solution. For compactness, we place the denominators accompanying the $l_{i}$ on the left hand side. Furthermore, in the expressions below, subscripts $\phi$ and $X$ denote differentiation with respect to $\phi$ and $X$, respectively,
\bea
\label{Eq: l1}
-d_{2}l_{1}&=&2A_{1}(2F_{1,\phi}+Q_{1})+2(2F_{1,\phi}+Q_{1})X(A_{4}+A_{5}X)+A_{2}(4F_{1,\phi}-2Q_{2}X)\nl+A_{3}X(4F_{1,\phi}+Q_{1}-Q_{2}X),
\eea
\bea
\label{Eq: l2}
F_{1}Xd_{2}l_{2}&=&-F_{1}X\big[-3A_{3}Q_{1}X+6A_{2}Q_{2}X+2A_{1}(Q_{1}+Q_{2}X)\big]-F_{1,\phi}X\bigg(X\big[4A_{1}^{2}\nl+4A_{1}X(A_{3}+A_{4}+A_{5}X)-3A_{3}(2F_{1}+A_{3}X^{2})\big]+4A_{2}\Big\{-3F_{1}+X\nl\times\big[4A_{1}+3X(A_{4}+A_{5}X)\big]\Big\}\bigg)+\tilde{f}\bigg(4F_{1}(Q_{1}+Q_{2}X)-2X\Big\{2A_{1}Q_{1}\nl+X\big[-2A_{2}Q_{2}+2Q_{1}(A_{4}+A_{5}X)+A_{3}(Q_{1}-Q_{2}X)\big]\Big\}\bigg),
\eea
\bea
\label{Eq: l4}
F_{1}Xd_{2}l_{4}&=&F_{1,\phi}X\Big\{4A_{1}^{2}+3X\big[-A_{3}^{2}X+4A_{2}(A_{4}+A_{5}X)\big]+4A_{1}\big[4A_{2}+X(A_{3}+A_{4}\nl+A_{5}X)\big]\Big\}+2\tilde{f}\Big\{2A_{1}Q_{1}+X\big[-2A_{2}Q_{2}+2Q_{1}(A_{4}+A_{5}X)+A_{3}(Q_{1}\nl-Q_{2}X)\big]\Big\},
\eea
\bea
\label{Eq: l5}
2d_{1}l_{5}=-2A_{1}+F_{2},
\eea
\bea
\label{Eq: l8}
d_{3}l_{8}=-8A_{1}F_{1}F_{1,X}+4A_{1}^{2}(\tilde{f}-2\tilde{f}_{X}X)+A_{4}^{2}X^{2}(-\tilde{f}+2\tilde{f}_{X}X),
\eea
\bea
\label{Eq: l9}
\frac{1}{2}X^{2}d_{1}d_{3}l_{9}&=&8A_{1}(\tilde{f}-F_{1})F_{1}(\tilde{f}+F_{1})+8A_{1}\tilde{f}\big[-2F_{1}\tilde{f}_{X}+(\tilde{f}+F_{1})F_{1,X}\big]X\nl+8A_{1}^{3}\tilde{f}_{X}X^{3}-2A_{1}A_{4}^{2}\tilde{f}_{X}X^{5}+A_{4}^{2}\tilde{f}X^{3}(\tilde{f}+F_{1}-2\tilde{f}_{X}X)+4A_{1}^{2}X\Big\{-\tilde{f}^{2}\nl+2F_{1}(F_{1}-2\tilde{f}_{X}X+F_{1,X}X)+\tilde{f}\big[F_{1}+2(\tilde{f}_{X}+F_{1,X})X\big]\Big\},
\eea
\bea
\label{Eq: l10}
2F_{1}Xd_{3}l_{10}&=&16A_{1}F_{1}^{3}+8\tilde{f}F_{1}(3A_{1}^{2}+A_{4}F_{1})X-8(2A_{1}^{2}+A_{4}F_{1})(2F_{1}\tilde{f}_{X}-\tilde{f}F_{1,X})X^{2}\nl-6A_{4}^{2}\tilde{f}F_{1}X^{3}+4A_{4}^{2}(2F_{1}\tilde{f}_{X}-\tilde{f}F_{1,X})X^{4},
\eea
\bea
\label{Eq: l11}
F_{1}Xd_{3}l_{11}=2(-4A_{1}^{2}+A_{4}^{2}X^{2})\big[-2F_{1}\tilde{f}_{X}X+\tilde{f}(F_{1}+F_{1,X}X)\big],
\eea
\bea
\label{Eq: l14}
-d_{1}d_{2}l_{14}&=&8A_{1}(A_{1}+2A_{2})+F_{1}+8\big[A_{2}A_{4}+A_{1}(A_{3}+A_{4})\big]X-2\big[A_{3}^{2}\nl-4(A_{1}+A_{2})A_{5}\big]X^{2},
\eea
\bea
\label{Eq: l15}
-Xd_{1}d_{2}l_{15}&=&8A_{1}^{2}(A_{1}+4A_{2})X^{2}+8A_{1}\big[3A_{2}A_{4}+A_{1}(A_{3}+A_{4})\big]X^{3}+2A_{1}\big[-3A_{3}^{2}\nl+4(A_{1}+3A_{2})A_{5}\big]X^{4}-8A_{1}F_{1}X(2A_{2}+A_{3}X)-4\tilde{f}\bigg(A_{3}X(2F_{1}\nl+A_{3}X^{2})+4A_{2}\Big\{F_{1}-X\big[A_{1}+X(A_{4}+A_{5}X)\big]\Big\}\bigg),
\eea
\bea
\label{Eq: l17}
-Xd_{1}d_{2}l_{17}&=&-8A_{1}^{3}X+6A_{1}A_{3}^{2}X^{3}+4\tilde{f}X\big[A_{3}^{2}X-4A_{2}(A_{4}+A_{5}X)\big]-8A_{1}A_{2}\big[2\tilde{f}\nl+3X^{2}(A_{4}+A_{5}X)\big]-8A_{1}^{2}X\big[4A_{2}+X(A_{3}+A_{4}+A_{5}X)\big],
\eea
\bea
\label{Eq: l18}
-\frac{1}{4}Xd_{1}d_{2}d_{3}l_{18}&=&8\tilde{f}^{3}(4A_{1}^{2}-A_{4}^{2}X^{2})\big[A_{1}+A_{2}+X(A_{3}+A_{4}+A_{5}X)\big]-8A_{1}F_{1}^{3}\big[4A_{1}^{2}\nl+8A_{1}A_{2}-A_{3}^{2}X^{2}+4A_{2}X(A_{4}+A_{5}X)+4A_{1}X(A_{3}+A_{4}+A_{5}X)\big]\nl+\tilde{f}^{2}\Big\{16A_{1}^{3}(A_{1}-2A_{2})X+16A_{1}^{2}\big[-3A_{2}A_{4}+A_{1}(A_{3}+A_{4})\big]X^{2}\nl+4A_{1}\big[3A_{1}A_{3}^{2}-A_{1}A_{4}^{2}+2A_{2}A_{4}^{2}+4A_{1}(A_{1}-3A_{2})A_{5}\big]X^{3}-4A_{4}^{2}\big[\nl-3A_{2}A_{4}+A_{1}(A_{3}+A_{4})\big]X^{4}-A_{4}^{2}\big[3A_{3}^{2}+4(A_{1}-3A_{2})A_{5}\big]X^{5}\nl-16\tilde{f}_{X}X(4A_{1}^{2}-A_{4}^{2}X^{2})\big[A_{1}+A_{2}+X(A_{3}+A_{4}+A_{5}X)\big]+8F_{1,X}(\nl-8A_{1}F_{1}+4A_{1}^{2}X-A_{4}^{2}X^{3})\big[A_{1}+A_{2}+X(A_{3}+A_{4}+A_{5}X)\big]\Big\}\nl+2A_{1}\tilde{f}_{X}X^{3}(4A_{1}^{2}-A_{4}^{2}X^{2})\Big\{4A_{1}^{2}+3X\big[-A_{3}^{2}X+4A_{2}(A_{4}+A_{5}X)\big]\nl+4A_{1}\big[4A_{2}+X(A_{3}+A_{4}+A_{5}X)\big]\Big\}+8A_{1}^{2}F_{1}F_{1,X}X\Big\{8A_{1}F_{1}+8A_{2}F_{1}\nl+4A_{1}^{2}X+8A_{3}F_{1}X-3A_{3}^{2}X^{3}+8F_{1}X(A_{4}+A_{5}X)+12A_{2}X^{2}(A_{4}\nl+A_{5}X)+4A_{1}X\big[4A_{2}+X(A_{3}+A_{4}+A_{5}X)\big]\Big\}+\tilde{f}\Bigg[-16A_{1}^{4}(A_{1}\nl+4A_{2})X^{2}-16A_{1}^{3}\big[3A_{2}A_{4}+A_{1}(A_{3}+A_{4})\big]X^{3}+4A_{1}^{2}\big[3A_{1}A_{3}^{2}\nl+A_{1}A_{4}^{2}+4A_{2}A_{4}^{2}-4A_{1}(A_{1}+3A_{2})A_{5}\big]X^{4}+4A_{1}A_{4}^{2}\big[3A_{2}A_{4}+A_{1}(A_{3}\nl+A_{4})\big]X^{5}+A_{1}A_{4}^{2}\big[-3A_{3}^{2}+4(A_{1}+3A_{2})A_{5}\big]X^{6}-F_{1}X(4A_{1}^{2}-A_{4}^{2}X^{2})\nl\times\big[4A_{1}^{2}+8A_{1}A_{2}-A_{3}^{2}X^{2}+4A_{2}X(A_{4}+A_{5}X)+4A_{1}X(A_{3}+A_{4}+A_{5}X)\big]\nl-2\tilde{f}_{X}X^{2}(4A_{1}^{2}-A_{4}^{2}X^{2})\Big\{4A_{1}^{2}+3X\big[A_{3}^{2}X-4A_{2}(A_{4}+A_{5}X)\big]+4A_{1}\big[\nl-2A_{2}+X(A_{3}+A_{4}+A_{5}X)\big]\Big\}+8A_{1}F_{1,X}\bigg(8A_{2}F_{1}^{2}+8(A_{3}+A_{4})F_{1}^{2}X\nl+4A_{1}^{3}X^{2}+4F_{1}(3A_{2}A_{4}+2A_{5}F_{1})X^{2}-3(A_{3}^{2}-4A_{2}A_{5})F_{1}X^{3}-A_{2}A_{4}^{2}X^{4}\nl-A_{4}^{2}(A_{3}+A_{4})X^{5}-A_{4}^{2}A_{5}X^{6}+A_{1}\Big\{8F_{1}^{2}-A_{4}^{2}X^{4}-4F_{1}X\big[-2A_{2}\nl+X(A_{3}+A_{4}+A_{5}X)\big]\Big\}+4A_{1}^{2}X\Big\{-F_{1}+X\big[A_{2}+X(A_{3}+A_{4}\nl+A_{5}X)\big]\Big\}\bigg)\Bigg],
\eea
\bea
\label{Eq: l19}
-\frac{1}{4}F_{1}X^{2}d_{1}d_{2}d_{3}l_{19}&=&32A_{1}^{2}F_{1}^{4}X(2A_{2}+A_{3}X)+16\tilde{f}^{3}X(F_{1}+F_{1,X}X)\big[A_{1}+A_{2}+X(A_{3}+A_{4}\nl+A_{5}X)\big]\big[4A_{1}^{2}+A_{4}(2F_{1}-A_{4}X^{2})\big]-8A_{1}^{2}F_{1}^{3}X^{2}\Big\{4A_{1}^{2}+3X\big[-A_{3}^{2}X\nl+4A_{2}(A_{4}+A_{5}X)\big]+4A_{1}\big[4A_{2}+X(A_{3}+A_{4}+A_{5}X)\big]\Big\}+4A_{1}F_{1}\tilde{f}_{X}X^{4}\nl\times\big[4A_{1}^{2}+A_{4}(2F_{1}-A_{4}X^{2})\big]\Big\{4A_{1}^{2}+3X\big[-A_{3}^{2}X+4A_{2}(A_{4}+A_{5}X)\big]\nl+4A_{1}\big[4A_{2}+X(A_{3}+A_{4}+A_{5}X)\big]\Big\}+16A_{1}^{2}F_{1}^{2} F_{1,X}X^{2}\bigg(X\big[4A_{1}^{2}\nl+4A_{1}X(A_{3}+A_{4}+A_{5}X)-3A_{3}(2F_{1}+A_{3}X^{2})\big]+4A_{2}\Big\{-3F_{1}\nl+X\big[4A_{1}+3X(A_{4}+A_{5}X)\big]\Big\}\bigg)+4\tilde{f}^{2}X\bigg(2F_{1}X(4A_{1}^{2}-A_{4}^{2}X^{2})\big[A_{1}^{2}\nl-3A_{1}A_{2}+A_{3}^{2}X^{2}-4A_{2}X(A_{4}+A_{5}X)+A_{1}X(A_{3}+A_{4}+A_{5}X)\big]\nl-8F_{1}\tilde{f}_{X}X\big[A_{1}+A_{2}+X(A_{3}+A_{4}+A_{5}X)\big]\big[4A_{1}^{2}+A_{4}(2F_{1}-A_{4}X^{2})\big]\nl+F_{1}^{2}\Big\{4A_{1}^{2}\big[2A_{2}+(A_{3}+A_{4})X\big]+4A_{1}A_{4}X\big[-2A_{2}+X(A_{3}+A_{4}\nl+A_{5}X)\big]-A_{4}X^{2}\big[A_{3}(-3A_{3}+A_{4})X+2A_{2}(7A_{4}+6A_{5}X)\big]\Big\}+F_{1,X}X\nl\times\Big\{16A_{1}^{4}X-4A_{1}^{2}(6A_{2}F_{1}+3A_{3}F_{1}X-A_{4}F_{1}X+A_{4}^{2}X^{3})+16A_{1}^{3}X\big[A_{2}\nl+X(A_{3}+A_{4}+A_{5}X)\big]+3A_{4}F_{1}X^{2}\big[A_{3}(A_{3}+A_{4})X-2A_{2}(A_{4}+2A_{5}X)\big]\nl-4A_{1}A_{4}X\big[X(A_{3}+A_{4}+A_{5}X)(-F_{1}+A_{4}X^{2})+A_{2}(2F_{1}+A_{4}X^{2})\big]\Big\}\bigg)\nl+\tilde{f}F_{1}\Bigg[32A_{1}F_{1}^{3}(2A_{2}+A_{3}X)-16A_{1}F_{1}^{2}X\Big\{4A_{1}A_{2}+X\big[-A_{3}^{2}X+4A_{2}\nl\times(A_{4}+A_{5}X)\big]\Big\}+4\tilde{f}_{X}X^{3}\big[-4A_{1}^{2}+A_{4}(-2F_{1}+A_{4}X^{2})\big]\Big\{4A_{1}^{2}+3X\nl\times\big[A_{3}^{2}X-4A_{2}(A_{4}+A_{5}X)\big]+4A_{1}\big[-2A_{2}+X(A_{3}+A_{4}+A_{5}X)\big]\Big\}\nl-3A_{1}X^{3}(4A_{1}^{2}-A_{4}^{2}X^{2})\Big\{4A_{1}^{2}+3X\big[-A_{3}^{2}X+4A_{2}(A_{4}+A_{5}X)\big]\nl+4A_{1}\big[4A_{2}+X(A_{3}+A_{4}+A_{5}X)\big]\Big\}-4A_{1}F_{1}X^{2}\Big\{-4A_{1}^{2}\big[2A_{2}+(A_{3}\nl-A_{4})X\big]+4A_{1}A_{4}X\big[4A_{2}+X(A_{3}+A_{4}+A_{5}X)\big]+A_{4}X^{2}\big[A_{3}(-3A_{3}\nl+A_{4})X+2A_{2}(7A_{4}+6A_{5}X)\big]\Big\}-4A_{1}F_{1,X}X\bigg(X\Big\{4A_{1}X(A_{3}+A_{4}\nl+A_{5}X)(-4F_{1}+A_{4}X^{2})+4A_{1}^{2}\big[-4F_{1}+(3A_{3}+A_{4})X^{2}\big]+3A_{3}\big[8F_{1}^{2}\nl+4A_{3}F_{1}X^{2}-A_{4}(A_{3}+A_{4}) X^{4}\big]\Big\}+2A_{2}\Big\{24F_{1}^{2}-8F_{1}X\big[4A_{1}+3X(A_{4}\nl+A_{5}X)\big]+X^{2}\big[12A_{1}^{2}+8A_{1}A_{4}X+3A_{4}X^{2}(A_{4}+2A_{5}X)\big]\Big\}\bigg)\Bigg],
\eea
\bea
\label{Eq: l21}
-\frac{1}{8}F_{1}X^{2}d_{1}d_{2}d_{3}l_{21}&=&8\tilde{f}^{3}(F_{1}+F_{1,X}X)(-4A_{1}^{2}+A_{4}^{2}X^{2})\big[A_{1}+A_{2}+X(A_{3}+A_{4}+A_{5}X)\big]\nl+4\tilde{f}^{2}X(-4A_{1}^{2}+A_{4}^{2}X^{2})\bigg(2A_{1}^{2}F_{1,X}X+F_{1}\Big\{-4A_{2}\tilde{f}_{X}-4\big[A_{2}A_{4}+(A_{3}\nl+A_{4})\tilde{f}_{X}\big]X+\big[A_{3}^{2}-4A_{5}(A_{2}+\tilde{f}_{X})\big]X^{2}\Big\}+2A_{1}\big[-2F_{1}(A_{2}+\tilde{f}_{X})\nl+A_{2}F_{1,X}X+(A_{3}+A_{4})F_{1,X}X^{2}+A_{5}F_{1,X}X^{3}\big]\bigg)-8A_{1}^{2}F_{1}^{2}F_{1,X}X^{2}\Big\{4A_{1}^{2}\nl+3X\big[-A_{3}^{2}X+4A_{2}(A_{4}+A_{5}X)\big]+4A_{1}\big[4A_{2}+X(A_{3}+A_{4}+A_{5}X)\big]\Big\}\nl-2A_{1}F_{1}\tilde{f}_{X}X^{3}(4A_{1}^{2}-A_{4}^{2}X^{2})\Big\{4A_{1}^{2}+3X\big[-A_{3}^{2}X+4A_{2}(A_{4}+A_{5}X)\big]\nl+4A_{1}\big[4A_{2}+X(A_{3}+A_{4}+A_{5}X)\big]\Big\}+\tilde{f}F_{1}\bigg(16A_{1}^{4}(A_{1}+4A_{2})X^{2}\nl+16A_{1}^{3}\big[3A_{2}A_{4}+A_{1}(A_{3}+A_{4})\big]X^{3}+4A_{1}^{2}\big[-3A_{1}A_{3}^{2}-A_{1}A_{4}^{2}-4A_{2}A_{4}^{2}\nl+4A_{1}(A_{1}+3A_{2})A_{5}\big]X^{4}-4A_{1}A_{4}^{2}\big[3A_{2}A_{4}+A_{1}(A_{3}+A_{4})\big]X^{5}-A_{1}A_{4}^{2}\big[\nl-3A_{3}^{2}+4(A_{1}+3A_{2})A_{5}\big]X^{6}+8A_{1}F_{1}^{2}\big[4A_{1}^{2}+8A_{1}A_{2}-A_{3}^{2}X^{2}+4A_{2}X\nl\times(A_{4}+A_{5}X)+4A_{1}X(A_{3}+A_{4}+A_{5}X)\big]+2\tilde{f}_{X}X^{2}(4A_{1}^{2}-A_{4}^{2}X^{2})\Big\{\nl4A_{1}^{2}+3X\big[A_{3}^{2}X-4A_{2}(A_{4}+A_{5}X)\big]+4A_{1}\big[-2A_{2}+X(A_{3}+A_{4}\nl+A_{5}X)\big]\Big\}-8A_{1}F_{1}F_{1,X}X\Big\{4A_{1}^{2}+3X\big[-A_{3}^{2}X+4A_{2}(A_{4}+A_{5}X)\big]\nl+4A_{1}\big[4A_{2}+X(A_{3}+A_{4}+A_{5}X)\big]\Big\}\bigg).
\eea
\section{Effective action coefficients}
\label{App: Effective action coefficients}
The coefficients of the action in Eq. \eqref{Eq: Effective action} are given below, where $\tilde{f}=F_{1}-\frac{1}{2}F_{2}X$. As in the previous appendix, we move the denominators to the left hand side to keep the expressions compact,
\bea
\label{Eq: Ptilde}
d_{2}\tilde{P}&=&4A_{1}\big[-(A_{1}+4A_{2})P+Q_{1}^{2}\big]X+\big[-4A_{1}(A_{3}+A_{4})P+3A_{4}(-4A_{2}P+Q_{1}^{2})\nl+2A_{1}Q_{1}Q_{2}\big]X^{2}+\big[3A_{3}^{2}P-4(A_{1}+3A_{2})A_{5}P+3A_{5}Q_{1}^{2}\nl-3A_{3}Q_{1}Q_{2}+(A_{1}+3A_{2})Q_{2}^{2}\big]X^{3}+2\tilde{f}\big[4(A_{1}+A_{2})P-Q_{1}^{2}+4(A_{3}+A_{4})PX\nl-2Q_{1}Q_{2}X+(4A_{5}P-Q_{2}^{2})X^{2}\big]+12F_{1,\phi}^{2}X\big[A_{1}+A_{2}+X(A_{3}+A_{4}+A_{5}X)\big]\nl+6F_{1,\phi}X\Big\{2A_{1}Q_{1}+X\big[-2A_{2}Q_{2}+2Q_{1}(A_{4}+A_{5}X)+A_{3}(Q_{1}-Q_{2}X)\big]\Big\}
\eea
\bea
\label{Eq: Q1tilde}
d_{2}\tilde{Q}_{1}&=&-16\tilde{f}\tilde{f}_{\phi}\big[A_{1}+A_{2}+X(A_{3}+A_{4}+A_{5}X)\big]+16F_{1}F_{1,\phi}\big[A_{1}+A_{2}+X(A_{3}+A_{4}+A_{5}X)\big]\nl+\tilde{f}_{\phi}X\Big\{8A_{1}^{2}+6X\big[-A_{3}^{2}X+4A_{2}(A_{4}+A_{5}X)\big]+8A_{1}\big[4A_{2}+X(A_{3}+A_{4}+A_{5}X)\big]\Big\}\nl+4F_{1}\Big\{2A_{1}Q_{1}+X\big[-2A_{2}Q_{2}+2Q_{1}(A_{4}+A_{5}X)+A_{3}(Q_{1}-Q_{2}X)\big]\Big\}
\eea
\bea
\label{Eq: Q2tilde}
-Xd_{2}\tilde{Q}_{2}&=&-16\tilde{f}\tilde{f}_{\phi}\big[A_{1}+A_{2}+X(A_{3}+A_{4}+A_{5}X)\big]+16F_{1,\phi}(F_{1}-3F_{1,X}X)\big[A_{1}+A_{2}\nl+X(A_{3}+A_{4}+A_{5}X)\big]+\tilde{f}_{\phi}X\Big\{8A_{1}^{2}+6X\big[-A_{3}^{2}X+4A_{2}(A_{4}+A_{5}X)\big]\nl+8A_{1}\big[4A_{2}+X(A_{3}+A_{4}+A_{5}X)\big]\Big\}+4F_{1}\Big\{2A_{1}Q_{1}+X\big[-2A_{2}Q_{2}\nl+2Q_{1}(A_{4}+A_{5}X)+A_{3}(Q_{1}-Q_{2}X)\big]\Big\}-12F_{1,X}X\Big\{2A_{1}Q_{1}\nl+X\big[-2A_{2}Q_{2}+2Q_{1}(A_{4}+A_{5}X)+A_{3}(Q_{1}-Q_{2}X)\big]\Big\}
\eea
\bea
\label{Eq: A1tilde}
2d_{1}\tilde{A}_{1}=-F_{2}^{2}X+2A_{1}(2F_{1}+F_{2}X)
\eea
\bea
\label{Eq: A2tilde}
\frac{1}{2}Xd_{1}d_{2}\tilde{A}_{2}&=&-8\tilde{f}^{3}\big[A_{1}+A_{2}+X(A_{3}+A_{4}+A_{5}X)\big]+F_{1}^{2}X\Big\{4A_{1}^{2}+4A_{1}X(A_{3}+A_{4}+A_{5}X)\nl+X\big[A_{3}^{2}X-4A_{2}(A_{4}+A_{5}X)\big]\Big\}-\tilde{f}^{2}X\Big\{4A_{1}^{2}+3X\big[A_{3}^{2}X\nl-4A_{2}(A_{4}+A_{5}X)\big]+4A_{1}\big[-2A_{2}+X(A_{3}+A_{4}+A_{5}X)\big]\Big\}\nl+\tilde{f}\bigg(4A_{1}^{3}X^{2}+8F_{1}^{2}\big[A_{2}+X(A_{3}+A_{4}+A_{5}X)\big]+4A_{1}^{2}X^{2}\big[4A_{2}\nl+X(A_{3}+A_{4}+A_{5}X)\big]+A_{1}\Big\{8F_{1}^{2}+3X^{3}\big[-A_{3}^{2}X+4A_{2}(A_{4}+A_{5}X)\big]\Big\}\bigg)
\eea
\bea
\label{Eq: A3tilde}
-\frac{1}{4}X^{2}d_{1}d_{2}\tilde{A}_{3}&=&\tilde{f}\bigg(8(A_{1}+A_{2})F_{1}^{2}+8F_{1}\big[(A_{3}+A_{4})F_{1}-2(A_{1}+A_{2})F_{1,X}\big]X+4\Big\{A_{1}^{3}\nl-4A_{1}A_{2}\tilde{f}_{X}+2A_{1}^{2}(2A_{2}+\tilde{f}_{X})+2F_{1}\big[A_{5}F_{1}-2(A_{3}+A_{4})F_{1,X}\big]\Big\}X^{2}\nl+4\big[3A_{1}A_{2}A_{4}+A_{1}^{2}(A_{3}+A_{4})-6A_{2}A_{4}\tilde{f}_{X}+2A_{1}(A_{3}+A_{4})\tilde{f}_{X}\nl-4A_{5}F_{1}F_{1,X}\big]X^{3}+\big[4A_{1}^{2}A_{5}+6(A_{3}^{2}-4A_{2}A_{5})\tilde{f}_{X}+A_{1}(-3A_{3}^{2}+12A_{2}A_{5}\nl+8A_{5}\tilde{f}_{X})\big]X^{4}\bigg)-8\tilde{f}^{3}\big[A_{1}+A_{2}+X(A_{3}+A_{4}+A_{5}X)\big]-16A_{1}F_{1}F_{1,X}X^{2}\big[A_{1}\nl+A_{2}+X(A_{3}+A_{4}+A_{5}X)\big]+F_{1}^{2}X\Big\{4A_{1}^{2}+4A_{1}X(A_{3}+A_{4}+A_{5}X)\nl+X\big[A_{3}^{2}X-4A_{2}(A_{4}+A_{5}X)\big]\Big\}+\tilde{f}^{2}X\Big\{-4A_{1}^{2}+8A_{1}A_{2}-4\big[-3A_{2}A_{4}\nl+A_{1}(A_{3}+A_{4})\big]X-\big[3A_{3}^{2}+4(A_{1}-3A_{2})A_{5}\big]X^{2}+16\tilde{f}_{X}\big[A_{1}+A_{2}+X(A_{3}\nl+A_{4}+A_{5}X)\big]\Big\}-2A_{1}\tilde{f}_{X}X^{3}\Big\{4A_{1}^{2}+3X\big[-A_{3}^{2}X+4A_{2}(A_{4}+A_{5}X)\big]\nl+4A_{1}\big[4A_{2}+X(A_{3}+A_{4}+A_{5}X)\big]\Big\}
\eea
\bea
\label{Eq: A4tilde}
-\frac{1}{8}X^{2}d_{1}d_{3}\tilde{A}_{4}&=&\bigg(8A_{1}^{3}\tilde{f}_{X}X^{3}(\tilde{f}-2\tilde{f}_{X}X)+A_{4}^{2}\tilde{f}X^{3}\big[F_{1}^{2}+2\tilde{f}_{X}X(-\tilde{f}+2\tilde{f}_{X}X)\big]\nl+4A_{1}^{2}X\Big\{2\tilde{f}^{2}\tilde{f}_{X}X-4F_{1}\tilde{f}_{X}X(F_{1}+2F_{1,X}X)+\tilde{f}\big[F_{1}^{2}+4F_{1}F_{1,X}X\nl+2(-2\tilde{f}_{X}^{2}+F_{1,X}^{2})X^{2}\big]\Big\}+2A_{1}\big[-4F_{1}^{4}+2A_{4}^{2}\tilde{f}_{X}^{2}X^{6}+4\tilde{f}^{2}(F_{1}\nl+F_{1,X}X)^{2}-\tilde{f}\tilde{f}_{X}X(8F_{1}^{2}+16F_{1}F_{1,X}X+A_{4}^{2}X^{4})\big]\bigg)
\eea
\bea
\label{Eq: A5tilde}
-X^{2}d_{1}d_{2}d_{3}\tilde{A}_{5}&=&1024A_{1}^{2}F_{1}^{3}F_{1,X}X\big[A_{1}+A_{2}+X(A_{3}+A_{4}+A_{5}X)\big]-1536A_{1}^{2}F_{1}^{2}F_{1,X}^{2}X^{2}\big[A_{1}\nl+A_{2}+X(A_{3}+A_{4}+A_{5}X)\big]+64\tilde{f}^{4}(4A_{1}^{2}-A_{4}^{2}X^{2})\big[A_{1}+A_{2}+X(A_{3}+A_{4}\nl+A_{5}X)\big]-64A_{1}F_{1}^{4}\big[4A_{1}^{2}+8A_{1}A_{2}-A_{3}^{2}X^{2}+4A_{2}X(A_{4}+A_{5}X)+4A_{1}X(A_{3}\nl+A_{4}+A_{5}X)\big]+8\tilde{f}^{3}\Big\{16A_{1}^{3}(A_{1}-2A_{2})X+16A_{1}^{2}\big[-3A_{2}A_{4}+A_{1}(A_{3}\nl+A_{4})\big]X^{2}+4A_{1}\big[3A_{1}A_{3}^{2}-A_{1}A_{4}^{2}+2A_{2}A_{4}^{2}+4A_{1}(A_{1}-3A_{2})A_{5}\big]X^{3}\nl-4A_{4}^{2}\big[-3A_{2}A_{4}+A_{1}(A_{3}+A_{4})\big]X^{4}-A_{4}^{2}\big[3A_{3}^{2}+4(A_{1}-3A_{2})A_{5}\big]X^{5}\nl-128A_{1}F_{1}F_{1,X}\big[A_{1}+A_{2}+X(A_{3}+A_{4}+A_{5}X)\big]-64A_{1}F_{1,X}^{2}X\big[A_{1}+A_{2}\nl+X(A_{3}+A_{4}+A_{5}X)\big]-32\tilde{f}_{X}X(4A_{1}^{2}-A_{4}^{2}X^{2})\big[A_{1}+A_{2}+X(A_{3}+A_{4}\nl+A_{5}X)\big]\Big\}-256A_{1}^{2}F_{1}\tilde{f}_{X}F_{1,X}X^{3}\Big\{4A_{1}^{2}+3X\big[-A_{3}^{2}X+4A_{2}(A_{4}+A_{5}X)\big]\nl+4A_{1}\big[4A_{2}+X(A_{3}+A_{4}+A_{5}X)\big]\Big\}-32A_{1}\tilde{f}_{X}^{2}X^{4}(4A_{1}^{2}-A_{4}^{2}X^{2})\Big\{4A_{1}^{2}\nl+3X\big[-A_{3}^{2}X+4A_{2}(A_{4}+A_{5}X)\big]+4A_{1}\big[4A_{2}+X(A_{3}+A_{4}+A_{5}X)\big]\Big\}\nl+8\tilde{f}^{2}X\Bigg[-16A_{1}^{4}(A_{1}+4A_{2})X-16A_{1}^{3}\big[3A_{2}A_{4}+A_{1}(A_{3}+A_{4})\big]X^{2}\nl+4A_{1}^{2}\big[3A_{1}A_{3}^{2}+A_{1}A_{4}^{2}+4A_{2}A_{4}^{2}-4A_{1}(A_{1}+3A_{2})A_{5}\big]X^{3}+4A_{1}A_{4}^{2}\big[3A_{2}A_{4}\nl+A_{1}(A_{3}+A_{4})\big]X^{4}+A_{1}A_{4}^{2}\big[-3A_{3}^{2}+4(A_{1}+3A_{2})A_{5}\big]X^{5}+32\tilde{f}_{X}^{2}X(4A_{1}^{2}\nl-A_{4}^{2}X^{2})\big[A_{1}+A_{2}+X(A_{3}+A_{4}+A_{5}X)\big]-16F_{1}F_{1,X}\Big\{-12A_{1}^{2}A_{2}\nl+A_{1}X\big[(3A_{3}^{2}+A_{4}^{2})X-12A_{2}(A_{4}+A_{5}X)\big]+A_{4}^{2}X^{2}\big[A_{2}+X(A_{3}+A_{4}\nl+A_{5}X)\big]\Big\}-8F_{1,X}^{2}X\Big\{16A_{1}^{3}-3A_{1}X\big[(-A_{3}^{2}+A_{4}^{2})X+4A_{2}(A_{4}+A_{5}X)\big]\nl-3A_{4}^{2}X^{2}\big[A_{2}+X(A_{3}+A_{4}+A_{5}X)\big]+4A_{1}^{2}\big[A_{2}+4X(A_{3}+A_{4}+A_{5}X)\big]\Big\}\nl-4\tilde{f}_{X}\bigg(16A_{1}^{4}X+8A_{1}A_{2}(-8F_{1}F_{1,X}+A_{4}^{2}X^{3})-4A_{1}X(A_{3}+A_{4}+A_{5}X)\nl\times(16F_{1}F_{1,X}+A_{4}^{2}X^{3})+3A_{4}^{2}X^{4}\big[-A_{3}^{2}X+4A_{2}(A_{4}+A_{5}X)\big]\nl+16A_{1}^{3}X\big[-2A_{2}+X(A_{3}+A_{4}+A_{5}X)\big]-4A_{1}^{2}\Big\{16F_{1}F_{1,X}+X^{2}\big[(-3A_{3}^{2}\nl+A_{4}^{2})X+12A_{2}(A_{4}+A_{5}X)\big]\Big\}\bigg)\Bigg]+8\tilde{f}\Bigg[-F_{1}^{2}X(4A_{1}^{2}-A_{4}^{2}X^{2})\big[4A_{1}^{2}\nl+8A_{1}A_{2}-A_{3}^{2}X^{2}+4A_{2}X(A_{4}+A_{5}X)+4A_{1}X(A_{3}+A_{4}+A_{5}X)\big]\nl+4\tilde{f}_{X}^{2}X^{3}(4A_{1}^{2}-A_{4}^{2}X^{2})\Big\{4A_{1}^{2}+3X\big[A_{3}^{2}X-4A_{2}(A_{4}+A_{5}X)\big]+4A_{1}\big[\nl-2A_{2}+X(A_{3}+A_{4}+A_{5}X)\big]\Big\}-8A_{1}F_{1,X}^{2}X\bigg(8A_{1}^{3}X^{2}-3(-8F_{1}^{2}+A_{4}^{2}X^{4})\nl\times\big[A_{2}+X(A_{3}+A_{4}+A_{5}X)\big]+4A_{1}^{2}X^{2}\big[-A_{2}+2X(A_{3}+A_{4}+A_{5}X)\big]\nl+3A_{1}\Big\{8F_{1}^{2}-X^{3}\big[4A_{2}A_{4}+(-A_{3}^{2}+A_{4}^{2}+4A_{2}A_{5})X\big]\Big\}\bigg)\nl+4A_{1}\tilde{f}_{X}X^{2}\bigg(16A_{1}^{4}X+3X(8F_{1}F_{1,X}+A_{4}^{2}X^{3})\big[A_{3}^{2}X-4A_{2}(A_{4}+A_{5}X)\big]\nl+16A_{1}^{3}X\big[4A_{2}+X(A_{3}+A_{4}+A_{5}X)\big]-4A_{1}\big[X(A_{3}+A_{4}+A_{5}X)(\nl-8F_{1}F_{1,X}+A_{4}^{2}X^{3})+4A_{2}(4F_{1}F_{1,X}+A_{4}^{2}X^{3})\big]+4A_{1}^{2}\Big\{8F_{1}F_{1,X}+X^{2}\big[\nl-(3A_{3}^{2}+A_{4}^{2})X+12A_{2}(A_{4}+A_{5}X)\big]\Big\}\bigg)+16A_{1}F_{1}F_{1,X}\bigg(8A_{1}^{3}X^{2}+(8F_{1}^{2}\nl-A_{4}^{2}X^{4})\big[A_{2}+X(A_{3}+A_{4}+A_{5}X)\big]+4A_{1}^{2}X^{2}\big[5A_{2}+2X(A_{3}+A_{4}\nl+A_{5}X)\big]+A_{1}\Big\{8F_{1}^{2}+X^{3}\big[-(3A_{3}^{2}+A_{4}^{2})X+12A_{2}(A_{4}+A_{5}X)\big]\Big\}\bigg)\Bigg]
\eea
\section{Class Ia coefficients}
\label{App: Class Ia coefficients}
In this appendix we present the explicit Class Ia coefficients $\tilde{p}$, $\tilde{q}_{1}$, $\tilde{q}_{2}$, $\tilde{a}_{1}$, $\tilde{a}_{2}$, $\tilde{a}_{3}$, $\tilde{a}_{4}$, and $\tilde{a}_{5}$ appearing in Eq. \eqref{Eq: Class Ia} for $P=Q_{1}=Q_{2}=0$, expressed in terms of the two functions $F_{1}(\phi,X)$ and $F_{2}(\phi,X)$ that define the metric-affine Class Ia sector,
\bea
\label{Eq: Class Ia coefficients}
&&\tilde{p}=-\frac{3F_{1,\phi}^{2}F_{2}X^{2}}{4(F_{1}^{2}+3F_{1,X}^{2}X^{2})},\nl
\tilde{q}_{1}=-2\tilde{f}_{\phi}-\frac{F_{1}F_{1,\phi}F_{2}X}{F_{1}^{2}+3F_{1,X}^{2}X^{2}},\nl
\tilde{q}_{2}=\frac{2\tilde{f}_{\phi}}{X}+\frac{F_{1,\phi}F_{2}(F_{1}-3F_{1,X}X)}{F_{1}^{2}+3F_{1,X}^{2}X^{2}},\nl
\tilde{a}_{1}=-\tilde{a}_{2}=-F_{2}+\frac{F_{1}}{X}+\frac{F_{1}^{2}F_{2}}{2(F_{1}^{2}+3F_{1,X}^{2}X^{2})},\nl
\tilde{a}_{3}=2F_{2,X}+\frac{(F_{1}-2F_{1,X}X)(2F_{1}^{2}+F_{1}F_{2}X+6F_{1,X}^{2}X^{2})}{X^{2}(F_{1}^{2}+3F_{1,X}^{2}X^{2})},\nl
\tilde{a}_{4}=-\frac{1}{9F_{1,X}^{2}X^{4}(F_{1}^{2}+3F_{1,X}^{2}X^{2})}\Big[2F_{1}^{5}-F_{1}^{4}(12F_{1,X}+F_{2})X-6F_{1}^{2}F_{1,X}^{2}X^{3}(12F_{1,X}-F_{2}\nl-3F_{2,X}X)-54F_{1,X}^{4}X^{5}(2F_{1,X}-F_{2,X}X)-6F_{1}^{3}F_{1,X}X^{2}(-8F_{1,X}+F_{2}+2F_{2,X}X)\nl-18F_{1}F_{1,X}^{3}X^{4}(-7F_{1,X}+F_{2}+2F_{2,X}X)\Big],\nl
\tilde{a}_{5}=\frac{F_{1}}{9F_{1,X}^{2}X^{5}(F_{1}^{2}+3F_{1,X}^{2}X^{2})}\Big[2F_{1}^{4}-F_{1}^{3}(12F_{1,X}+F_{2})X+6F_{1}F_{1,X}^{2}(-6F_{1,X}+F_{2})X^{3}\nl+6F_{1}^{2}F_{1,X}X^{2}(5F_{1,X}-F_{2}-2F_{2,X}X)+36F_{1,X}^{3}X^{4}(2F_{1,X}-F_{2,X}X)\Big].
\eea
\bibliographystyle{JHEP}
\bibliography{References}
\end{document}